\documentclass[pra,11pt,onecolumn, showkeys,showpacs]{revtex4}

\usepackage{times}

\begin{document}

\title{Eavesdropping on the improved three-party quantum secret sharing protocol }

\author{Gan Gao\footnote{Corresponding author. E-mail:
gaogan0556@163.com}}

\affiliation{Department of Electrical Engineering, Tongling
University, Tongling 244000, China}

\date{\today}

\begin{abstract}

Lin et al. [S. Lin, F. Gao, Q.-y. Wen, F.-c. Zhu, Opt. Commun.
\textbf{281} (2008) 4553] pointed that the multiparty quantum secret
sharing protocol [Z.-j. Zhang, G. Gao, X. Wang, L.-f. Han, S.-h.
Shi, Opt. Commun.  \textbf{269}  (2007) 418] is insecure and
proposed an improved three-party quantum secret sharing protocol. In
this paper, we study the security of the improved three-party
quantum secret sharing protocol and find that it is still insecure.
Finally, a further improved three-party quantum secret sharing
protocol is proposed.
\end{abstract}



\maketitle

\section{Introduction}

In the last decade, as the principles of quantum mechanics are
introduced, a lot of interesting applications are realized in the
field of cryptography. Quantum secret sharing (QSS) [1-25] is one of
these applications, which allows that a secret message is spitted
into several pieces by a boss, and each agent owns a piece, and no
subset of agents can be sufficient to extract the boss's secret
message, but the whole set can. The first QSS protocol in which a
three-particle entangled Greenberger-Horne-Zeilinger (GHZ) state was
used is proposed by Hillery {\it et al.} \cite{1}. Although this
protocol has elegantly shown the essence of QSS, it is hard to
realize experimentally because of the inefficiency as regards the
generation of a three-particle entangled state. Recently, in order
to increase the practical feasibility, Zhang {\it et al.} [15]
utilized Einstein-Podolsky-Rosen (EPR) pairs and the five local
operations to propose a novel QSS protocol. However, it is a slight
pity that this protocol has a drawback of security, which was
pointed out by Lin {\it et al.} [16]. Lin {\it et al.} showed that
the last agent may solely obtain half of Alice's secret messages
without the other agents' helps, moreover, they gave an improvement
of Zhang {\it et al.} QSS protocol. Later on, Wang {\it et al.} [17]
claimed that the three-party case in Lin {\it et al.} improved
protocol [16] is secure, and pointed out that the $n$-party
($n\geq4$) case is insecure, and they show that the first agent and
the last agent in the improved QSS protocol [16] may collaborate to
eavesdrop Alice's secret messages without introducing any error.
Obviously, both Lin {\it et al.} attack and Wang {\it et al.} attack
are from the inside dishonest agents in QSS, and the attack power of
the dishonest agent is very strong because he (she) has a chance to
tell a lie during the checking security. Only if the lie is
successfully constructed, he (she) can eavesdrop the secret messages
without being detected by the other participants. Therefore, we
should mainly focus on the dishonest agent's attack while analyzing
the security of the QSS protocol.

In this paper, we study the security of the three-party case of Lin
{\it et al.} improved QSS protocol [16] and find that it is also
insecure. Before describing our attack strategy, first, let us
review Lin {\it et al.} improved three-party protocol [16] as
follows: (1)Bob prepares photons $h$ and $t$ in one of four Bell
states:
$\psi_{ht}^{\pm}=(|0\rangle|1\rangle\pm|1\rangle|0\rangle)_{ht}/\sqrt{2}$,
$\phi_{ht}^{\pm}=(|0\rangle|0\rangle\pm|1\rangle|1\rangle)_{ht}/\sqrt{2}$.
Then he sends photon $t$ to Charlie and retains photon $h$ in his
site. (2)After receiving photon $t$, firstly, Charlie ascertains
whether photon $t$ is a single photon [18]. If not, the
communication will be terminated. Otherwise, he performs one of the
five local operations: $I$, $\sigma_{x}$, $\sigma_{y}$,
$\sigma_{z}$, $H$ on photon $t$. The probabilities that the five
operations are selected by her are 1/8, 1/8, 1/8, 1/8 and 1/2,
respectively. Here, $I=|0\rangle\langle0|+|1\rangle\langle1|$,
$\sigma_{z}=|0\rangle\langle0|-|1\rangle\langle1|$,
$\sigma_{y}=|0\rangle\langle1|-|1\rangle\langle0|$,
$\sigma_{x}=|0\rangle\langle1|+|1\rangle\langle0|$, $H=
(|0\rangle\langle0|-|1\rangle\langle1|+|0\rangle\langle1|+|1\rangle\langle0|)/\sqrt{2}$.
Then he sends photon $t$ to Alice. (3)After receiving photon $t$,
Alice randomly switches between the control mode and the message
mode. In the control mode, Alice randomly selects one action from
the two choices: One is that she lets Bob use
$\{|0\rangle,|1\rangle\}$ or
$\{|h\rangle=(|0\rangle+|1\rangle)/\sqrt{2},|v\rangle=(|0\rangle-|1\rangle)/\sqrt{2}\}$
to measure photon $h$, and tell her his measurement outcome and
initial Bell state. Then Alice requires Charlie to announce their
operations. The other is that Alice first lets Charlie announce
their operations, and then asks Bob to perform a measurement on
photon $h$ and tell her his measurement outcome and initial Bell
state. Next, Alice uses correct measuring basis to measure the $t$
photon. By comparing her measurement outcome with the deduced
outcome, Alice can judge whether quantum channel is secure. If the
quantum channel is attacked, the communication is aborted.
Otherwise, the transmission goes on to Step (1). In the message
mode, Alice performs a unitary operation ($I$, $\sigma_{x}$,
$\sigma_{y}$, $\sigma_{z}$) on photon $t$ to encode her secret
messages. After her encoding, Alice sends all the $t$-photons of
message mode as a sequence ($t$-sequence) to Bob in one
communication. Before sending $t$-sequence, Alice prepares a certain
number of single photons (checking photons) randomly in one of four
states: $|0\rangle$, $|1\rangle$, $|h\rangle$, $|v\rangle$, and
inserts these checking photons into $t$-sequence. And then, she
sends $t$-sequence to Bob. (4)After Bob receives the sequence, Alice
tells Bob the positions of checking photons in the $t$-sequence and
the initial states of all checking photons. Bob picks out the
checking photons and uses the suitable basis to measure them. And
then, by comparing his measurement outcomes with the initial states,
Bob can judges whether quantum channel between Alice and him is
secure. After confirming that no eavesdropping exists, they can
extract Alice's secret messages if Bob and Charlie collaborate.
(5)At last, Alice announces a small part of the secret messages so
that Bob and Charlie can execute the message authentication process.

\section{Security leak of Lin et al. improved three-party QSS
protocol}

We can see that, in Lin {\it et al .} improved three-party protocol
[16], Alice inserts the checking photons into the $t$-sequence in
order to prevent Charlie from eavesdropping. By making a
single-photon measurement on each checking photon, Alice may judge
whether the quantum channel between Alice and Bob is safe. Though
the process of the security-check is added, their improved
three-party quantum secret sharing protocol is still insecure. In
what follows, we will detailedly analyse why it isn't secure.

In advance, Bob prepares two EPR photon pairs. Suppose that one pair
is in $\psi_{ab}^{-}$ and the other $\phi_{ht}^{+}$. According to
Step (1), Bob sends photon $t$ to Charlie. After receiving photon
$t$, Charlie performs one of the five local operations: $I$,
$\sigma_{x}$, $\sigma_{y}$, $\sigma_{z}$, $H$ on it, and then sends
photon $t$ to Alice. When photon $t$ is traveling between Charlie
and Alice, Bob intercepts it, and stores it well. At the same time,
Bob sends photon $b$ (that is from $\psi_{ab}^{-}$), instead of
photon $t$, to Alice. Alice randomly switches the control mode and
the message mode. In the control mode, when Alice requires Bob to
make single-photon measurement on photon $h$, firstly, Bob
immediately makes Bell state measurement on photons $a$ and $h$.
Obviously, this is the entangle swapping process of Bell states.
Suppose that the operation performed by Charlie is $H$, the whole
system state can be written as follows:
\begin{equation}
\psi^{-}_{ab} \otimes\phi^{+}_{ht}\rightarrow\psi^{-}_{ab}H
\phi^{+}_{ht}=\frac{1}{2}[\phi^{+}_{ah}(\psi^{+}_{bt}-\phi^{-}_{bt})
+\phi^{-}_{ah}(\phi^{+}_{bt}-\psi^{-}_{bt})-\psi^{+}_{ah}(\phi^{+}_{bt}+\psi^{-}_{bt})+\psi^{-}_{ah}(\phi^{-}_{bt}+\psi^{+}_{bt})]
\end{equation}
Suppose that Bob's Bell state measurement outcome on photons $a$ and
$h$ is $\psi^{+}_{ah}$. According to Equation (1), photons $b$ and
$t$ must be in $(\phi^{+}_{bt}+\psi^{-}_{bt})$. Next, Bob makes a
comparison for $\psi^{+}_{ah}$ and $\psi_{ab}^{-}$, and gets the
unitary operation $\sigma_{z}$. This kind of Bell state comparison
method and its comparison steps can be consulted in the paper [26].
By the way, here the four operations $I$, $\sigma_{x}$,
$\sigma_{y}$, $\sigma_{z}$ are similar to the four operators
$U_{1}$, $U_{2}$, $U_{3}$, $U_{4}$ in the paper [26]. After making
Bell state measurement, Bob uses $\{|0\rangle,|1\rangle\}$ or
$\{|h\rangle,|v\rangle\}$ to measure photon $t$ and tells Alice his
single-qubit measurement outcome. According to Step (3), Bob still
needs to tell Alice his initial Bell state. In order that his
replacing trick is not detected by Alice, Bob may not directly say
that the initial Bell state is $\phi_{ht}^{+}$, but should tell the
lie that it is $\sigma_{z}\phi_{ht}^{+}=\phi_{ht}^{-}$ ($\sigma_{z}$
is just the gotten operator that Bob makes the Bell state
comparison). So his replacing action will not be detected by Alice.
Here, we can't help asking why Bob's replacing action isn't
detected? Continuing to analyse, the key of the question will be
obtained. We know, that Alice deduces which state her and Bob's hand
photons are in depends on the information published by Bob and
Charlie. Only if it is satisfied that Alice's deducing state and the
$(\phi^{+}_{bt}+\psi^{-}_{bt})$ are the same states, Bob's replacing
trick can not be detected by Alice. From the above content, we can
see that the initial state published by Bob is $\phi_{ht}^{-}$ and
the operation published by Charlie is $H$, so Alice can only deduce
as follows:
\begin{equation}
\phi_{ht}^{-}\longrightarrow
H\phi_{ht}^{-}=\phi^{+}_{ht}+\psi^{-}_{ht}
\end{equation}
Indeed, the $(\phi^{+}_{ht}+\psi^{-}_{ht})$ and the
$(\phi^{+}_{bt}+\psi^{-}_{bt})$ are the same states. Hence, Bob's
replacing trick does not introduce any error. As for the other Bell
states that Bob's measurement outcomes on photons $a$ and $h$ are,
the law exists {\it as of old}. Here, for saving space of a whole
page, we don't list out the others' deducing processes again. So Bob
may evade Alice's security-check successfully in the case that
Charlie's operation is $H$. If Charlie's operation is one of four
unitary operations: $I$, $\sigma_{x}$, $\sigma_{y}$, $\sigma_{z}$,
is Bob's replacing trick also feasible? The answer is "yes". Suppose
that the two Bell states prepared by Bob are still $\psi_{ab}^{-}$
and $\phi_{ht}^{+}$, and the operation performed by Charlie is $I$.
So the whole system state may be written as follows:
\begin{equation}
\psi^{-}_{ab} \otimes  \phi^{+}_{ht}\rightarrow\psi^{-}_{ab} \otimes
I\phi^{+}_{ht}= \frac{1}{2} ( - \phi^{+}_{ah}\psi^{-}_{bt} +
\phi^{-}_{ah}\psi^{+}_{bt}   -
 \psi^{+}_{ah}\phi^{-}_{bt}  +  \psi^{-}_{ah}\phi^{+}_{bt} )
\end{equation}
Assume that Bob's measurement outcome on photons $a$ and $h$ is
$\phi^{-}_{ah}$. According to Equation (3), photons $b$ and $t$ must
be in $\psi^{+}_{bt}$. Similarly, Bob gets the unitary operation
$\sigma_{x}$ by comparing $\phi^{-}_{ah}$ with $\psi^{-}_{ab}$ [26].
So, he may tell the lie that it is
$\sigma_{x}\phi_{ht}^{+}=\psi_{ht}^{+}$ when Alice requires him to
publish the initial Bell state. According to Bob's $\psi_{ht}^{+}$
and Charlie's $I$, Alice deduces that
$I\psi_{ht}^{+}=\psi_{ht}^{+}$. It is evident that the
$\psi^{+}_{ht}$ and the $\psi^{+}_{bt}$ are the same states. So Bob
is also able to do his replacing trick in the case that Charlie's
operation is one of four unitary operations. In a word, no matter
what Charlie's operation is, Bob's replacing trick is not detected
by Alice in the control mode. When the message mode is switched
into, since Alice doesn't know that Bob has done the replacing trick
and regards photon $b$ as photon $t$ {\it as of old}, she encodes
her secret messages by performing a unitary operation on photon $b$.
Then she sends photon $b$ back to Bob. Bob very easily gets Alice's
secret messages by making Bell state measurement on photons $a$ and
$b$ without Charlie's helps.

So far, we have successfully proposed a attack for Lin {\it et al .}
improved three-party secret sharing protocol [16]. Meanwhile, we
have also proved Wang {\it et al.} statement that the three-party
case of Lin {\it et al .} improved protocol is secure is not
correct. Obviously, that our attack and Wang {\it et al.} attack
[17] are put together shows that Lin {\it et al .} improved QSS
protocol [16] is completely insecure. Next, let us discuss how to
further modify Lin {\it et al .} improved three-party QSS protocol
so that it can resist this kind of attack. For the integrity, we
describe the modified Lin {\it et al .} improved three-party
protocol as follows in brief.

($1'$) Bob prepares a batch of EPR photon pairs, which each pair is
randomly in one of four Bell states. He takes out photon $t$ from
each EPR pair to form one sequence, called $t$ sequence. The partner
photon $h$ in one EPR pair forms another sequence, called $h$
sequence. Then Bob sends the $t$ sequence to Charlie.

($2'$) After receiving the $t$ sequence, firstly, Charlie ascertains
whether the photons in $t$ sequence are a single photon [18]. If the
multi-photon signal is detected, the communication will be
terminated. Otherwise, he performs one of the five local operations:
$I$, $\sigma_{x}$, $\sigma_{y}$, $\sigma_{z}$ and $H$ on each photon
in the t sequence. The probabilities that the five operations are
selected by him are 1/8, 1/8, 1/8, 1/8 and 1/2, respectively. Next,
Charlie prepares a certain number of decoy photons, which each decoy
photon is randomly in one of four states: $|0\rangle, |1\rangle,
|h\rangle, |v\rangle$. He inserts these decoy photons into the $t$
sequence, and then sends the $t$ sequence to Alice.

($3'$) After confirming that Alice receives the $t$ sequence,
Charlie tells Alice the position of each decoy photon in the $t$
sequence and the state of each decoy photon. Afterwards, Alice uses
the appropriate measuring basis to measure each decoy photon. So she
can judge whether the quantum channel between hers and Charlie is
secure. Next, what Alice still needs to do is the same as that in
Step (3).

Steps ($4'$), ($5'$) are same to Steps (4), (5) in Lin {\it et al .}
improved three-party QSS protocol.

Up to now, we have proposed a further improved three-party QSS
protocol (Note that we only give the further improved three-party
QSS case here). In contrast to Lin {\it et al .} improved protocol
[16], our further improved protocol only adds one security-check
process between Charlie and Alice. By the way, this process is also
realized by utilizing some photons randomly in four states:
$|0\rangle, |1\rangle, |h\rangle, |v\rangle$. If Bob uses the above
replacing trick to attack this further improved protocol, his
eavesdropping action will fail because this process exists. The
reason is that, firstly, Bob doesn't know the positions of decoy
photons in $t$ sequence and the states of decoy photons, secondly,
both the decoy photon and the photon $t$ in $t$ sequence are in
maximally mixed state
$\rho=\frac{1}{2}(|0\rangle\langle0|+|1\rangle\langle1|)$ for Bob so
that he can't distinguish them, when he replaces the $t$ sequence
with another sequence prepared by him, Bob inevitably introduces
some errors to the decoy photons. As a result, his replacing trick
will be detected. In other words, our further improved three-party
QSS protocol can stand against the above proposed attack.

In the end, we discuss the securities of channels in this further
improved three-party QSS protocol one by one. From Steps ($2'$) and
($3'$), we see that the checking photons are used to assure the
security of the channel between Alice and Bob, and the decoy photons
the security of the channel between Charlie and Alice. Moreover, the
decoy photons and the checking photons are same and the purposes
using them are also same, that is, to analyse whether the
eavesdropping exists. As a matter of fact, the procedures that
utilize decoy photons and checking photons to check eavesdropping
are equivalent to the security checking in BB84 protocol [27]. Since
BB84 protocol has been proved to be unconditionally secure
[28,29,30], both of the Alice-Bob channel and the Charlie-Alice
channel are also safe in this further improved protocol. Now, there
leaves only the Bob-Charlie channel whose security is not discussed.
If we regard Bob as one party, and regard two of Charlie and Alice
as the other party, the security checking procedure for the
Bob-Charlie channel is essentially the same as that in BBM92
protocol [31]. Similarly, since BBM92 protocol has been proved to be
secure [32,33], his eavesdropping would be detected if the outside
eavesdropper attacks the Bob-Charlie channel. On the basis of these
analysis, this further improved three-party QSS protocol is secure.

\section{Summary}

We have shown that, by doing the replacing trick, Bob is able to
solely eavesdrop Alice's secret messages without introducing any
error in Lin {\it et al .} improved three-party QSS protocol. That
is, an efficient attack has been proposed for Lin {\it et al.}
improved three-party QSS protocol by us. Obviously, the trait of
this attack is that the Bell state comparison and the entanglement
swapping are employed. By the way, the Bell state comparison [26]
has been also used in the proposed attack strategy [25]. In
addition, after giving this attack, we further modify Lin {\it et al
.} improved three-party protocol so that it can stand against this
attack. Finally, by means of analyzing each channel, we discuss the
security of this further improved three-party QSS protocol.

\noindent {\bf Note added}$-$Recently, we detect that the proposed
attack in this paper has a small drawback. The drawback mainly
focuses on that the correlation of the two particles in one Bell
state are not always identical after the local operation $H$ is
performed on one of the two particles. For example,
$H\psi_{bt}^{+}=(|0\rangle|-\rangle+|1\rangle|+\rangle)_{bt}/\sqrt{2}=(|0\rangle|+\rangle-|1\rangle|-\rangle)_{tb}/\sqrt{2}$.
This means the proposed attack needs to be further optimized.\\


\begin{thebibliography}{99}



\bibitem{1} M. Hillery, V. Buzk and A. Berthiaume, Phys. Rev. A
$59$ (1999) 1829.

\bibitem{2}  D. Gottesman,  Phys. Rev. A $61$ (1999) 042311.

\bibitem{3}  R. Cleve, D. Gottesman and H. K. Lo,  Phys. Rev. Lett.
$83$ (1999) 648.

\bibitem{4}  V. Karimipour and A. Bahraminasab,  Phys. Rev. A $65$
(2000) 042320.

\bibitem{5}  W. Tittel, H. Zbinden and N. Gisin,  Phys. Rev. A $63$
(2001) 042301.

\bibitem{6}  S. Bandyopadhyay,  Phys. Rev. A  $62$ (2000) 012308.

\bibitem{7}  C. P. Yang and J. Gea-Banacloche, J. Opt. B: Qantum
Semiclass. Opt. $3$ (2001) 407.

\bibitem{8}  H. F. Chau,  Phys. Rev. A $66$ (2002) 060302.

\bibitem{9}  L. Y. Hsu,  Phys. Rev. A $68$ (2003) 022306.

\bibitem{11}  L. Xiao, G. L. Long, F. G. Deng and J. W. Pan,  Phys.
Rev. A $69$ (2004) 052307.

\bibitem{11}   Y. Q. Zhang, X. R. Jin and S. Zhang,  Phys. Lett. A
$341$ (2005) 380.

\bibitem{12}  A. M. Lance, T. Symul, W. P. Bowen {\it et al.},
Phys. Rev. Lett. $92$ (2004) 177903.

\bibitem{13} C. Schmid, P. Trojek, M. Bourennane {\it et al.},
Phys. Rev. Lett. $95$ (2005) 230505.


\bibitem{14}  G. P. He, Phys. Rev. Lett. $98$ (2007) 028901.

\bibitem{15} Z. J. Zhang, G. Gao, X. Wang, L. F. Han, S. H. Shi,
Opt. Commun. $269$ (2007) 418.

\bibitem{16} S. Lin, F. Gao, Q. Y. Wen, F. C. Zhu, Opt. Commun.
$281$ (2008) 4553.

\bibitem{17} T. Y. Wang, Q. Y. Wen, F. Gao, S. Lin, F. C. Zhu,
Phys. Lett. A $373$ (2008)65.

\bibitem{18}  F. G. Deng, X. H. Li, H. Y. Zhou and Z. J. Zhang,
Phys. Rev. A $72$ (2005) 044302.

\bibitem{19} I. C. Yu, F. L. Lin and C. Y. Huang,  Phys. Rev. A
$78$ (2008) 012344.

\bibitem{20}  S. Gaertner, C. Kurtsiefer, M. Bourennane, and H.
Weinfurter1, Phys. Rev. Lett. $98$ (2007) 020503.



\bibitem{21}  F. G. Deng, X. H. Li , H. Y. Zhou, Phys. Lett. A $372$
(2008)1957.

\bibitem{22} D. Markham, B. C. Sanders, Phys. Rev. A $78$ (2008)
042309.

\bibitem{23} J. Bogdanski, N. Rafiei, and M. Bourennane, Phys. Rev.
A $78$ (2008) 062307.

\bibitem{24}  G. Gao, Commun. Theor. Phys $52$ (2009) 421.

\bibitem{25}  G. Gao, Opt. Commun. $282$ (2009) 4464.

\bibitem{26}  G. Gao, Opt. Commun. $281$ (2008) 876.



\bibitem{27}  C. H. Bennett, G. Brassard, in a {\it Proceedings of
the IEEE International Conference on Computers, Systems and Signal
Processings, Bangalore, India} (IEEE, New York, 1984) p175.


\bibitem{28}  D. Mayers, in {\it Advances in Cryptology-Proceedings
of Crypto ¡¯96} (Springer-Verlag, New York, 1996), p. 343.

\bibitem{29}  E. Biham, M. Boyer, P. O. Boykin, T. Mor, and V.
Roychowdhury, in {\it Proceedings of the Thirty-Second Annual ACM
Symposium on Theory of Computing} (ACM Press, New York, 2000), p.
715.

\bibitem{30} P. W. Shor and J. Preskill, Phys. Rev. Lett. $85$
(2000) 441.


\bibitem{31}  C. H. Bennett, G. Brassard, N. D. Mermin, Phys. Rev.
Lett. $68$ (1992) 557.

\bibitem{32} H. Inamori, L. Rallan, and V. Vedral, J. Phys. A:
Math. Gen. $34$ (2001) 6913.

\bibitem{33} E. Waks, A. Zeevi, and Y. Yamamoto, Phys. Rev. A $65$
(2002) 052310.


\end{thebibliography}
\enddocument